\documentclass[aps,prl,reprint,groupedaddress]{revtex4-1}

\begin{document}

\title{Comment on ``Magnetic Levitation Stabilized by Streaming Fluid Flows''}

\author{Andrey Kuznetsov}
\email[]{a.v.kuznetsov@bk.ru}

\affiliation{Independent researcher, Irkutsk, Russia}

\maketitle
The authors of the Letter \cite{PhysRevLett.121.064502} discovered an interesting phenomenon. However, the explanation of radial stability of levitation \cite{PhysRevLett.121.064502} is not completely satisfactory: ``In experiments, we observe that the flea is unable to stay centered above the drive magnet below a critical viscosity. Computationally we observe that a simple numerical model of an unconstrained flea that excludes fluid inertia is also radially unstable. Both of these observations suggest a complex hydrodynamic origin to the radial stability.'' I (1) realized stable levitation of the same type in water and (2) found that simple model can demonstrate radial stabilization of a flea without taking into account fluid flows. Thus, the conclusion about basic origin of stabilization should be reconsidered.

Take a certain example of the model demonstrating radial stabilization. Both the flea and the drive magnet are described as magnetic dipoles of lengths $L_f$ and $L_d$, composed of point monopoles of strengths $\pm q_f$ and $\pm q_d$, respectively. Magnetic force between monopoles is determined by magnetic Coulomb's law $|{\bf F_C}|=\mu_0 q_f q_d / 4 \pi r^2$, where $\mu_0$ is the vacuum permeability. Although this model is physically incorrect, it can be used to describe the interaction of thin magnets \cite{MagneticPole}. Friction and gravity forces, ${\bf F_f}=-k {\bf v}$ and ${\bf F_g}=m {\bf g}$, are applied to the flea monopoles. Here $k$ is friction coefficient, ${\bf v}$ is velocity, $m$ is mass, and ${\bf g}$ is free-fall acceleration. The drive magnet rotates with a frequency $\nu_d$. Initially, the flea is placed horizontally at a height $h_0$ above the drive magnet with some radial displacement $d<(L_d-L_f)/2$. Numerical simulations show radial stabilization of the flea, for example, with the following values: $L_f=12 \, \mathrm{mm}$,  $L_d=25 \, \mathrm{mm}$, $\nu_d=50 \, \mathrm{s^{-1}} $, $\mu_0 q_f q_d  / 4 \pi=5\times10^{-5} \, \mathrm{N \, m^{2}}$, $m=3 \, \mathrm{g}$, $k=0.3 \, \mathrm{N \, s \, m^{-1}}$, $g=9.8 \, \mathrm{m \, s^{-2}}$, $h_0=20 \, \mathrm{mm}$, and $d=4 \, \mathrm{mm}$ \footnote{See supplementary video at https://youtu.be/3BVp3SOXvWY}. The origin of stabilization here is a centripetal component of vibrational repulsive force, related to radial component of the flea oscillations. This centripetal force can appear if $L_f<L_d$. A friction is necessary for damping only.

Levitation in water was realized with experimental setup built on the basis of Intllab stirrer MS-500 \cite{Note1}. Its motor shaft was extended up to 13 cm to reduce orientational action of magnetic field of the motor (it turned out that the flea is more sensitive to external fields in water than in high viscosity fluids). A stack of neodymium disc magnets of 5 mm in diameter was used as a flea. The drive magnet was comprised of two disc magnets of 15 mm in diameter with center-to-center distance of 25 mm. The optimal length of the stack was found to be about $12 \, \mathrm{mm}$. The flea was submerged in water with a cotton pad.

In addition, it was found that the flea can levitate in air for about 1.5 minutes \cite{Note1}. The flea was more sensitive to external fields in air than in water. It was oriented even by Earth's field. The presence of centripetal component of vibrational force was evident, since the flea oscillated in the vicinity of the drive magnet vertical axis. However, owing to insufficient damping, an amplitude of oscillations of the flea gradually increased, leading to its flying out.

Conclusion: the basic origin of radial stability is a centripetal component of the same vibrational force that makes a flea levitate.

Recommended reading on vibrational forces: \cite{feynman1963feynman, landau1976mechanics, blekhman}.

Acknowledgments: thanks to the team of educational project GetAClass for their inspirational video on the effect of magnetic stirrer bar levitation; special thanks to Andrey Schetnikov for useful discussions.

\end{document}